\documentclass[manuscript]{acmart}

\usepackage{wrapfig}
\usepackage{tikz}
\usetikzlibrary{trees}
\tikzstyle{every node}=[draw=black,thick,anchor=west]
\tikzstyle{schema}=[draw=black,fill=gray!30, very thick]
\tikzstyle{table}=[dashed,fill=gray!15]

\setcopyright{acmcopyright}
\copyrightyear{2020}
\acmYear{2020}
\acmDOI{}

\acmConference[]{SIGMOD Blog: \url{https://wp.sigmod.org/?p=3138}}{December, 2020}{}
\begin{document}
%\title{CogMatch: Cognitive Bias Aware Matching via Crowdsourcing}

%\title{InCogn: Incorporating Cognitive Biases in Crowdsourced Matching}
%\title{Endorsing Cognitive Biases for Crowdsourced Matching}
%\title{InCognito: Introducing Cognitive Biases to Crowdsourced Matching}
%\title{\textsc{InCognitoMatch}: Introducing Cognitive Biases to Crowdsourced Matching}
\title{Human's Role in-the-Loop}
%CaMaCrowd

\author{Avigdor Gal, Roee Shraga}
\affiliation{%
	\institution{Technion -- Israel Institute of Technology}
	\city{Haifa}
	\country{Israel}
}
\email{avigal@technion.ac.il, shraga89@campus.technion.ac.il}
%\and
%\author{Roee Shraga}
%\affiliation{%
%	\institution{Technion -- Israel Institute of Technology}
%	\city{Haifa}
%	\country{Israel}
%}
%\email{shraga89@campus.technion.ac.il}

%	\begin{abstract}
%abstract
%	\end{abstract}
\maketitle

	\section{Introduction}\label{sec:intro}
%The research of data integration spans over multiple decades, holds both theoretical and practical appeal, and enjoys a continued interest by researchers and practitioners. Efforts in the area include, among other things, process matching, schema matching, and entity resolution. Matching problems have been historically treated as semi-automated tasks in which correspondences are generated by automatic algorithms and subsequently validated by a single human expert. The reason for that is the inherent assumption that humans ``do it better." 

The research of data integration spans over multiple decades, holds both theoretical and practical appeal, and enjoys a continued interest by researchers and practitioners. Efforts in the area include, among other things, process matching, schema matching, and entity resolution. Matching problems have been historically treated as semi-automated tasks in which correspondences are generated by automatic algorithms and subsequently validated by human expert(s). The reason for that is the inherent assumption that humans ``do it better." 

%In recent years, there has been an evolution of data accumulation, management, analytics, and visualization (also known as \emph{big data}). Big data is primarily about collecting volumes of data in an increased velocity from various sources and varying veracity. Big data is characterized by technological advancements such as Internet of things (accumulation), cloud computing (management), and deep learning (analytics). Putting it all together provides us with a new, exciting, and challenging research agenda.

Data integration has been recently challenged by the need to handle large volumes of data, arriving at high velocity from a variety of sources, which demonstrate varying levels of veracity. This challenging setting, often referred to as {\em big data}, renders many of the existing techniques, especially those that are human-intensive, obsolete. Big data also produces technological advancements such as Internet of things, cloud computing, and deep learning, and accordingly, provides a new, exciting, and challenging research agenda. 

Given the availability of data and the improvement of machine learning techniques, this blog discusses the respective roles of humans and machines in achieving cognitive tasks in matching, aiming to determine whether traditional roles of humans and machines are subject to change
~\cite{roee2018sigmod,shraga2020phdworkshop}. 
Such investigation, we believe, will pave a way to better utilize both human and machine resources in new and innovative manners. 

We shall discuss two possible modes of change, namely {\em humans out} and \emph{humans in}. \emph{Humans out} aim at exploring out-of-the-box latent matching reasoning using machine learning algorithms when attempting to overpower human matcher performance. Pursuing out-of-the-box thinking, machine and deep learning can be involved in matching. \emph{Humans in} explores how to better involve humans in the matching loop by assigning human matchers with a symmetric role to algorithmic matcher in the matching process.

We start by scoping the discussion and introduce the matching problem. Then, we separate the discussion to the role of humans in-the-loop to humans out and humans in.

\section{Matching 101}
Modern industrial and business processes require intensive use of large-scale data integration techniques to combine data from multiple heterogeneous data sources into meaningful and valuable information. Such integration is performed on structured and semi-structured datasets from various sources such as SQL and XML schemata, entity-relationship diagrams, ontology descriptions, Web service specifications, interface definitions, process models, and Web forms. Data integration plays a key role in a variety of domains, including data warehouse loading and exchange, data wrangling,
%~\cite{kandel2011wrangler}, 
data lakes, aligning ontologies for the Semantic Web, Web service composition,
%~\cite{lemos2016web}, 
and business document format merging ({\em e.g.}, orders and invoices in e-commence).
%~\cite{RAHM2001}. As an example, a shopping comparison app that supports queries such as ``the cheapest computer among retailers'' or ``the best medical specialist for Crohn's disease in Crete'' requires integrating and matching several data sources of product purchase orders and medical records.

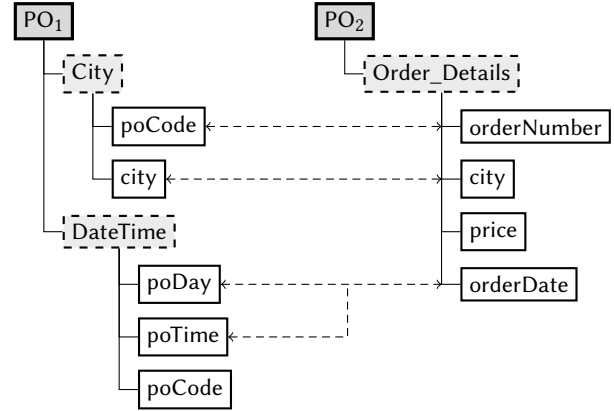
\begin{wrapfigure}[14]{R}{0.5\textwidth}
%	\begin{center}
%		\includegraphics[width=0.5\columnwidth]{./images/S1_S2_small}
%	\end{center}
%	\caption{Schema Matching Example.}
%	\label{ex1}
%\end{wrapfigure}
%\begin{figure}[h]  
	\centering
	\begin{tikzpicture}[%
		grow via three points={one child at (0.25,-0.7) and
			two children at (0.25,-0.7) and (0.25,-1.4)},
		edge from parent path={(\tikzparentnode.south) |- (\tikzchildnode.west)}]
		\node [schema] (a) {\sf PO$_{1}$}
		child { node [table] {\sf City}
			child { node (pocode) {\sf poCode}}
			child { node (pocity) {\sf city}}}
		child [missing] {}
		child [missing] {}
		child { node[table] {\sf DateTime}
			child { node (poday) {\sf poDay}}
			child { node (potime) {\sf poTime}}
			child { node {\sf poCode}}
		}
		;
		
		%grow via three points={one child at (0.25,-1) and
		% two children at (0.25,-5.5) and (0.25,-2.5)}
		\node [schema, right of=a,node distance=4cm] {\sf PO$_{2}$}
		child { node [table] {\sf Order\_Details}
			child { node (ordernumber) {\sf orderNumber}}
			child { node (ordercity) {\sf city}}
			child { node (orderprice){\sf price}}
			child { node (orderdate){\sf orderDate}}}
		;
		\draw[<->, densely dashed] (pocode.east) -- ([xshift=-.25cm]ordernumber.west);
		\draw[<->, densely dashed] (pocity.east) -- ([xshift=-.25cm]ordercity.west);
		\draw[<->, densely dashed] (poday.east) -- ([xshift=-.25cm]orderdate.west);
		\draw[<-, densely dashed] (potime.east) -- ([xshift=.5cm]potime.east)  -| ([xshift=-1.5cm]orderdate.west);
	\end{tikzpicture}
	\caption{Schema Matching Example.}
	\label{ex1}
\end{wrapfigure}  

A major challenge in data integration is a matching task, which creates correspondences between model elements, may they be schema attributes (see, for example, Figure~\ref{ex1}), ontology concepts, model entities, or process activities. Matching research has been a focus for multiple disciplines including Databases,
%~\cite{RAHM2001},
 Artificial Intelligence,
 %~\cite{de2018machine}, 
 Semantic Web,
 %~\cite{EUZENAT2007a}, 
 Process Management,
 %~\cite{leopold2012probabilistic}, 
 and Data Mining.
 %~\cite{lrsmTech}. 
 Most studies have focused on designing high quality matchers, automatic tools for identifying correspondences. Several heuristic attempts %({\em e.g.}, COMA~\cite{DO2002a}) 
 were followed by theoretical grounding, % ({\em e.g.}, see~\cite{BELLAHSENE2011,GAL2011}). 
showing that matching is inherently an uncertain decision making process due to ambiguity and heterogeneity %of data description concepts. The challenge of matching schemata lies in the heterogeneity
 of structure, semantics, and forms of representation of identical concepts.

\begin{wrapfigure}[11]{R}{0.5\textwidth}
	\begin{center}
		\includegraphics[width=0.5\columnwidth]{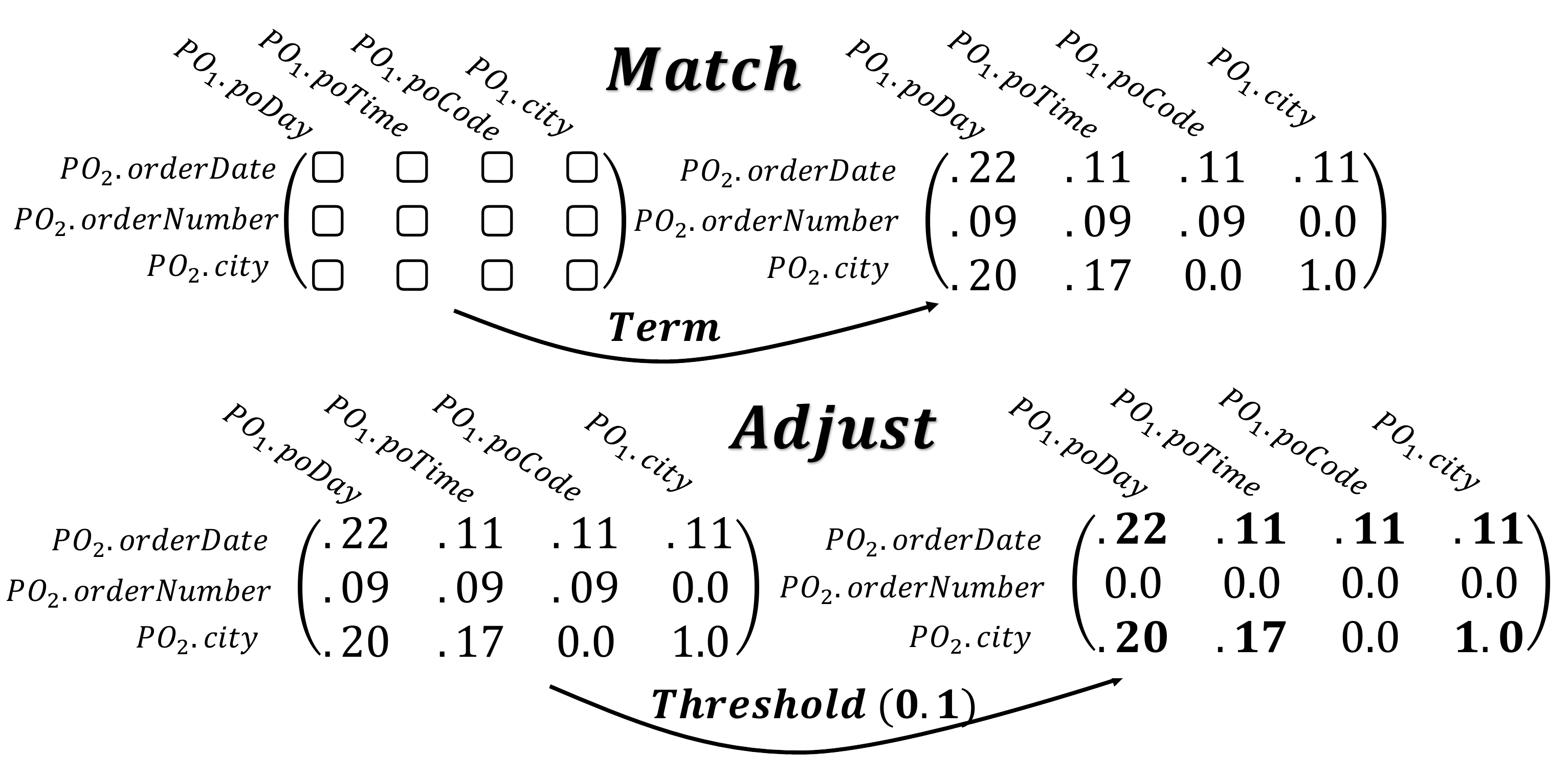}
	\end{center}
	\caption{Similarity Matrix Adjustment Illustration.}
	\label{ex11}
\end{wrapfigure}

We position the matching task as a {\em similarity matrix adjustment process} (see Figure~\ref{ex11} for an illustration). A matcher's output is conceptualized as a (possibly high dimensional) similarity matrix, in which a matcher records the similarity of elements from multiple data sources. Such elements can be attributes in a relational schema, tuples in databases that possibly represent the same real-world entity, {\em etc.} A major benefit of this approach is the ability to view schema matching as a machine learning task, paving the way to utilize deep learning models and achieve significant improvements in the ability to correctly match real-world data.

%One of the benefits of doing so is the enabling of a view of schema matching as a machine learning task, leading to a natural formulation using deep learning models that can support a significant improvement in the ability to correctly match real-world schemata.

The quality of automatic matching outcome is typically assessed using some evaluation metric ({\em e.g.,} Precision, Recall, F1). Applying such metrics usually requires human involvement to validate the decisions made by automatic matchers.~\cite{zhang2018reducing}.
 Yet, human validation of matching requires domain expertise~\cite{dragisic2016user} 
 and may be laborious,~\cite{zhang2018reducing}, 
 biased,~\cite{ackerman2019cognitive}, 
 and diverse.~\cite{Ross2010,HILDA18}. 
 This, in turn, limits the amount of qualitative labels that can be provided for supervised learning, especially when new domains are introduced. Matching predictors %~\cite{MCD,Sagi2013} 
 have been proposed in the literature as alternative evaluators for matching outcome, opting to correlate well with evaluation metrics created from human judgment. %Previous works have been focused so far on manually designed features and their combination~\cite{roee2018icdm}. Furthermore, trying to ``adjust'' (improve) schema matching outcome, previous works have utilized several human crafted rules and heuristics~\cite{chen2018biggorilla,do2007matching,fernandez2018seeping}.

It becomes clear that while matching tasks have been largely dependent on human validation, human performance in matching and validation does not meet the expectations of those who would like matches with high confidence. For many years there was no other option rather than trusting humans and secretly hoping our experts indeed perform well. The introduction of effective machine longer offers a promise of a better matching world, as will be discussed next.  

\section{Humans Out}\label{sec:out}

The \emph{Humans Out} approach seeks matching subtasks, traditionally considered to require cognitive effort, in which humans can be excluded. An initial good place to start is with the basic task of identifying correspondences. We note that many contemporary matching algorithms use heuristics, where each heuristic applies some semantic cue to justify an alignment between elements. For example, string-based matchers use string similarity as a cue for item alignment. We observe that such heuristics, in essence, encode human intuition about matching. Research shows that human matching choices can be reasonably predicted by classifying them into types, where a type corresponds to an existing heuristic. Therefore, we can argue that some of the cognitive effort of human matchers can be replaced by automated solutions. We next briefly describe two examples aiming to enhance the automation of matching using machine learning.

\textbf{Learning to Rerank Matches:} Choosing the best match from a top-$K$ is basically a cognitive task, usually done by humans. A recent work offers an algorithmic replacement to humans in selecting the best match~\cite{gal2019learning,roee2018icdm}, a task traditionally reserved for human verifiers. The novelty of this work is in the use of similarity matrices as a basis for learning features, creating feature-rich datasets that fit learning and enriches algorithmic matching beyond that of human matching. The suggested ranking framework adopts a learning-to-rank approach, utilizing matching predictors as features. An interesting aspect of this work is a bound on the size of $K$, given a desired level of confidence in finding the best match, justified theoretically and validated empirically. This bound is useful for top-$K$ algorithms and, as psychological literature suggests~\cite{schwartz2004paradox}, also applicable when introducing a list of options to humans.

\begin{wrapfigure}[12]{L}{0.5\textwidth}
	\centering
%	\begin{center}
		\includegraphics[width=0.48\columnwidth]{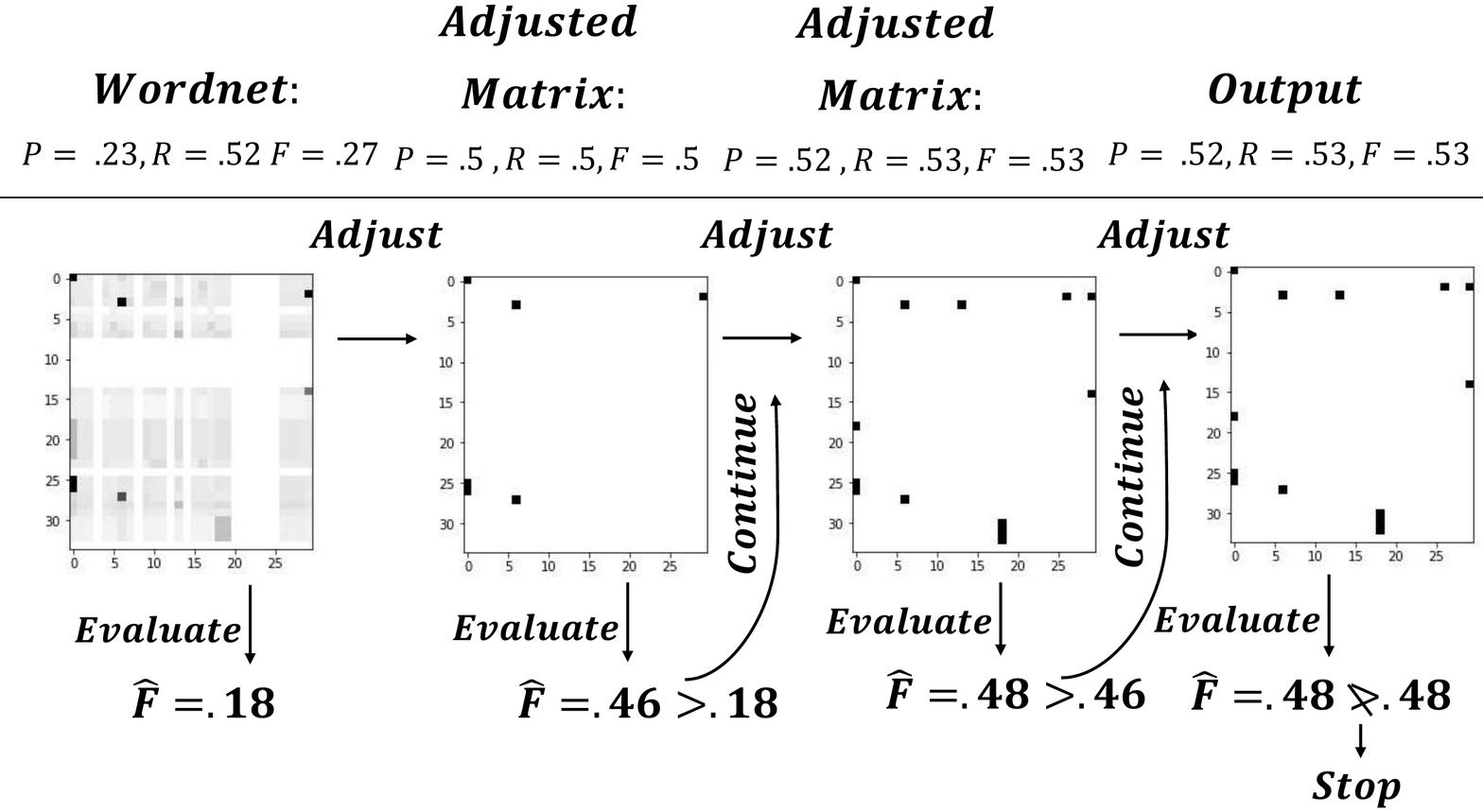}
%	\end{center}
	\vskip-.55in
	\caption{Deep Similarity Matrix Adjustment and Evaluation Example.}
	\label{ex2}
\end{wrapfigure}

\textbf{Deep Similarity Matrix Adjustment and Evaluation:} Traditionally, the adjustment and evaluation of matching results were done heuristically or by humans. A recent paper shows that these two processes can benefit from deep learning techniques~\cite{shraga2020}. The work offers a novel post processing step for schema matching that manipulates, using deep neural networks, similarity matrices, created by state-of-the-art algorithmic matchers (see Figure~\ref{ex2} for example). The suggested methodology provides a data-driven approach for extracting hidden representative features for an automatic schema matching process, removing the requirement for manual feature engineering. Moreover, it enhances the ability to introduce new data sources to existing systems without the need to rely on either domain experts (knowledgeable of the domain but less so on the best matchers to use) or data integration specialists (who lack sufficient domain knowledge).

%An additional example of a cognitive task involves choosing the best match from a top-$K$ list of matches, which can be transformed into re-ranking of top-$K$ matches so that the best match is ranked at the top. The literature offers an algorithm that show good results when tested on real-world as well as synthetic datasets, offering a replacement to humans in selecting the best match, a task traditionally reserved for human verifiers.	
%The novelty of this line of work is in the use of similarity matrices as a basis for learning features, creating feature-rich datasets that fit learning and offers a feature aggregation that is needed to enrich algorithmic matching beyond that of human matching. To create a reranking framework, it was suggested to adopt a learning-to-rank approach,%h~\cite{lambdamart}, 
%utilizing previously suggested matching predictors%~\cite{Prediction,MCD}
% as features. An interesting aspect of is a bound on the size of $K$, given a desired level of confidence in finding the best match, justified theoretically and validated empirically. This bound is useful for top-$K$ algorithms%~\cite{Macdonald2013} 
% and, as psychological literature suggests, also applicable when introducing a list of options (as in the traditional top-$K$ setting) to humans.%~\cite{schwartz2004paradox}. In addition, we provide a set of novel features to complement state-of-the-art and show further improvement in matching and ranking quality.

	\section{Humans In}\label{sec:in}

The \emph{Humans In} approach aims at investigating whether the current role humans take in the matching process is effective (spoiler: it is not) and whether an alternative role can improve overall performance of the matching process (spoiler: it can). A recent study, aided by metacognitive models, analyzes the consistency of human matchers~\cite{ackerman2019cognitive}. Three main consistency dimensions were explored as potential cognitive biases, taking into account the time it takes to reach a matching decision, the extent of agreement among human matchers and the assistance of algorithmic matchers. In particular, it was shown that when an algorithmic suggestion is available, humans tend to accept it to be true, in sharp contradiction to the conventional validation role of human matchers. 

\begin{wrapfigure}[12]{R}{0.5\textwidth}
	\begin{center}
		\includegraphics[width=0.5\columnwidth]{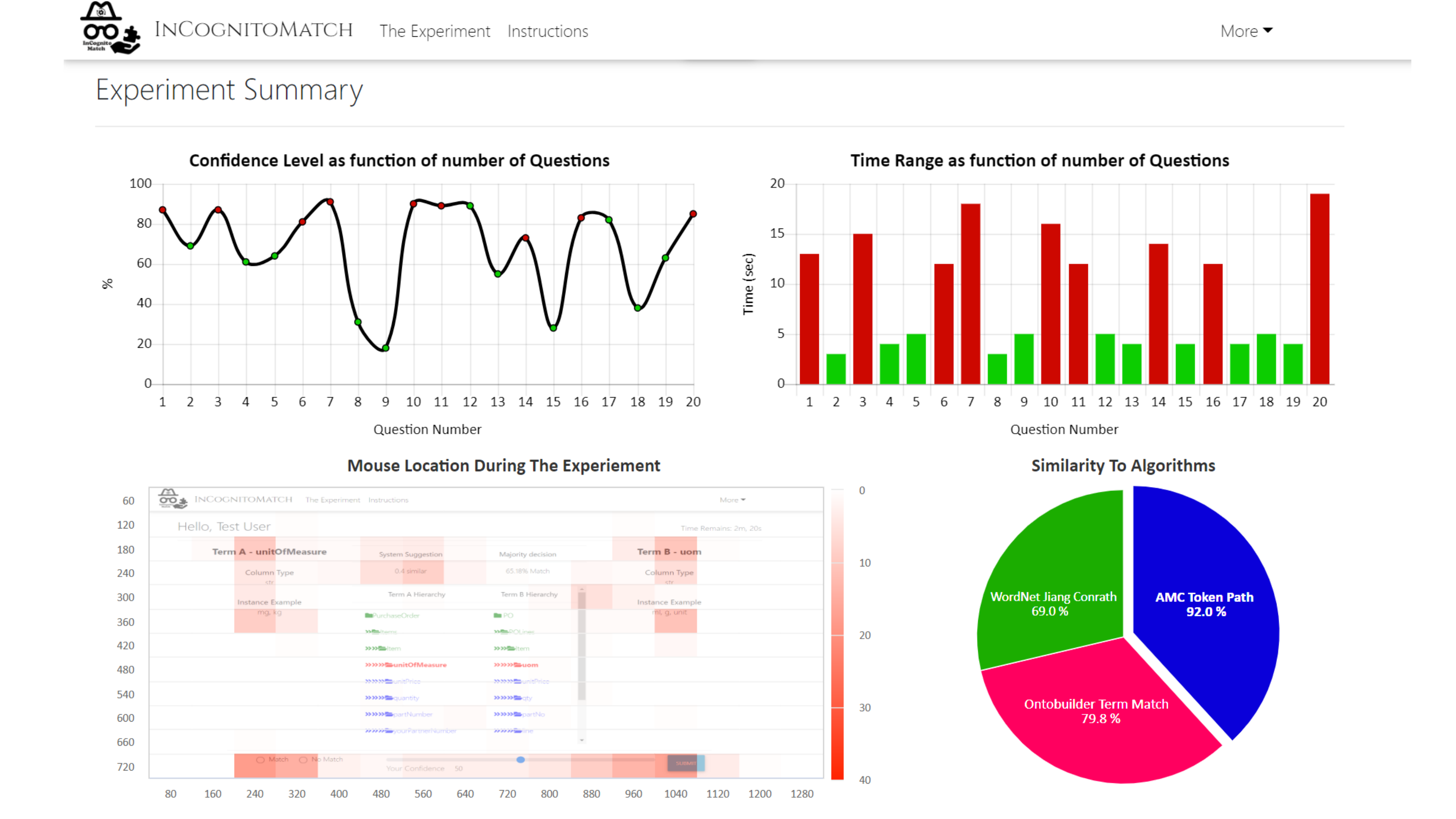}
	\end{center}
	\caption{Incognitomatch User Statistics.}
	\label{ex3}
\end{wrapfigure}

All these dimensions were found predictive of both confidence and accuracy of human matchers. Specifically, these biases can be used to recognize and capture predictive behaviors in the human decision making process (\emph{e.g.,} decision times and screen scrolls by mouse tracking). A recent research provides a machine learning framework to \emph{characterize domain experts} that can be used to identify reliable and valuable human experts, those humans whose proposed correspondences can mostly be trusted to be valid~\cite{roee2021icde,shraga2021jdiq}.

A new matching crowdsourcing application (see Figure~\ref{ex3} for an example screenshot), called Incognitomatch~\cite{roee2020sigmod}\footnote{\url{https://agp.iem.technion.ac.il/incognitomatch/}}, was launched to support the gathering of behavioral information and offer support in analyzing human matching behavior.

%The \emph{Humans In} approach aims at investigating whether the current role humans take in the matching process is effective (spoiler: it is not) and whether an alternative role can improve overall performance of the matching process (spoiler: it can). A recent study,%~\cite{humanMatching}, 
%aided by metacognitive models, analyzes the consistency of human matchers. Three main consistency dimensions were explored as potential cognitive biases, taking into account the time it takes to reach a matching decision, the extent of agreement among human matchers and the assistance of algorithmic matchers. In particular, it was shown that when an algorithmic suggestion is available, humans tend to accept it to be true, in sharp contradiction to the conventional validation role of human matchers.

%	Interestingly enough, all these dimensions were found predictive of both confidence and accuracy of human matchers. This indicates that 1) humans have cognitive biases affecting their ability to provide consistent matching decisions, and 2) that such biases has predictive value in determining to what extent a human matcher's alignment decision is accurate. 

	\section{Conclusion}\label{sec:con}
	
	In this blog we presented our approach for human involvement in the matching loop, introducing tasks where humans can be replaced and emphasizing our vision for understanding human behavior to allow better engagement. An additional overarching goal is to propose a common matching framework that would allow treating matching as a unified problem whether we match schemata attributes, ontology elements, process activities, entity's tuples, {\em etc.} 
	
	We envision future research to rely on machine and deep learning to generate new and innovative methods to improve matching problems, making extensive use of similarity matrices. Using machine learning for this purpose immediately raises the issue of shortage of labeled data to offer supervised learning. Hence, pursuing less-than-supervised ({\em e.g.}, unsupervised, weakly supervised) methods would be a natural next step to pursue.
	
%	 Future research will recruit the nest of machine learning (f0r now it is deep learning) to identify methods to improve matching problems such as process and schema matching, making extensive use of similarity matrices. We believe that applying non-linear transformations over a similarity matrix can reveal hidden relationships between elements that can be used for adjusting a given similarity matrix to better fit a matching problem.
%	 
%	 We also promote research that aims at predicting humans qualification to serve as ``experts'' for a matching task. Exploring predictive behaviors that capture the process of human matching by transforming physical aspects (such as time, screen scrolls, mouse tracking, and eye movement) can turn them into features that, in turn, can be used for examining the role of humans in the matching process. One possible outcome could allow matching systems to carefully select a matching expert that fits a task.
%
%	An overarching goal is to propose a common matching framework that would allow treating matching as a unified problem whether we match process activities, schemata attributes, entity's tuples, {\em etc.} Using machine learning for this purpose immediately raises the issue of shortage of labeled data to offer supervised learning. %~\cite{lrsmTech,mudgal2018deep,jabeen2017make}. 
%	Hence, pursuing less-than-supervised ({\em e.g.}, unsupervised, weakly supervised) methods would be a natural next step to pursue.

\section*{About the Authors}
 Avigdor Gal is a Professor of Data Science at the Technion -- Israel Institute of Technology. He is also the Academic Director of Data Science \& Engineering for the university’s undergraduate and graduate programs, a member of the management team of the Technion's Data Science Initiative and the head of the Big Data Integration laboratory in the Technion. In the current age of big data, his research is focused on developing novel models and algorithms for data integration broadly construed, and in particular the investigation of aspects of uncertainty when integrating data from multiple data sources. He is an expert in information systems and data science, and more recently, in machine learning and artificial intelligence.

Roee Shraga is a Postdoctoral fellow at the Technion – Israel Institute of Technology, from which he received a PhD degree in 2020 in the area of Data Science. Roee has published more than a dozen papers in leading  journals and conferences on the topics of data integration, human-in-the-loop, machine learning, process mining, and information retrieval. He is also a recipient of several PhD fellowships including the Leonard and Diane Sherman Interdisciplinary Fellowship (2017), the Daniel Excellence Scholarship (2019), and the Miriam and Aaron Gutwirth Memorial Fellowship (2020). 

\section*{Acknowledgments}
We would like thank Prof. Rakefet Ackerman, Dr. Haggai Roitman, Dr. Tomer Sagi, Dr. Ofra Amir,and Coral Scharf for their involvement in our research.

%    \balance
	\bibliographystyle{ACM-Reference-Format}
	\bibliography{ltsLong}
\end{document}